\documentclass[aps,twocolumn,showpacs]{revtex4}
\usepackage{epsfig}

\usepackage{graphicx}% Include figure files
\usepackage{dcolumn}% Align table columns on decimal point

\begin{document}

\title{Anomalous quantum chaotic behavior in nanoelectromechanical structures}
\author{Luis G. C. Rego$^1$, Andre Gusso$^2$, and M. G. E. da Luz$^2$}
\affiliation{1-Departamento de F\'{\i}sica,
Universidade Federal de Santa Catarina, Florian\'opolis, SC,
88040-900, Brazil, \\
2-Departamento de F\'{\i}sica, Universidade Federal do Paran\'a,
Curitiba, PR, 81531-990, Brazil}

\date{\today}

\begin{abstract}
It is predicted that for sufficiently strong electron-phonon coupling
an anomalous quantum chaotic behavior develops in certain types of
suspended electro-mechanical nanostructures, here represented by a thin
cylindrical quantum dot (billiard) on a suspended rectangular dielectric
plate.
The deformation potential and piezoelectric interactions are considered.
As a result of the electron-phonon coupling between the two systems the
spectral statistics of the electro-mechanic eigenenergies exhibit an
anomalous behavior.
If the center of the quantum dot is located at one of the symmetry axes
of the rectangular plate, the energy level distributions correspond to
the Gaussian Orthogonal Ensemble (GOE), otherwise they belong to the
Gaussian Unitary Ensemble (GUE), even though the system is time-reversal
invariant.
\end{abstract}
\pacs{85.85.+j, 05.45.Mt, 73.21.-b}

\maketitle

%----------------
%INTRODUCTION
%----------------

The possibility of engineering devices at the nano and micro scales
has openned a great avenue for testing fundamental aspects of quantum
theory, otherwise difficult to probe in natural atomic size systems.
In particular, mesoscopic structures have played an important role in
the experimental study of quantum chaos \cite{Stockmann}, mainly
through the investigation of the transport properties of quantum dots
\cite{Marcus} and quantum well structures \cite{Fromhold} in the
presence of magnetic field.
However, some hard to control characteristics of such structures can
prevent the full observation of quantum chaotic behavior.
For instance, the incoherent influence of the bulk on the electronic
dynamics hinders the observation of the so called eigenstate scars
\cite{Heller} in quantum corrals \cite{Crommie}.
Also, Random Matrix Theory (RMT) predictions to the Coulomb blockade
peaks in quantum dots may fail due to coupling to the environment
\cite{Madger}.
Alternatively, suspended nanostructures, due to their particular
architecture, are ideal candidates for investigating as well as
implementing coherent phenomena in semiconductor devices
\cite{nano,Blencowe}.
In special, to understand in a controlled way how phonons influence
electronic states and affect the system dynamics, possibly leading to a
chaotic behavior.
Such point is of practical relevance since it bears the question of
stability of quantum computers \cite{Chuang,Cleland}, whose actual
implementation could be prevented by the emergence of chaos \cite{Georgeot}.

A remarkable characteristic of the quantum chaotic systems is the manifestation of universal
statistical features that occur irrespective to their physical nature (e.g., the energy spectra of
spinless particles). According to RMT the resulting spectra for systems with and without time-reversal
invariance (TRI) are typically described by random matrices of the Gaussian Orthogonal Ensemble (GOE)
and Gaussian Unitary Ensemble (GUE), respectively. This property was conjectured by Bohigas et al.
\cite{Bohigas} and has been firmly established by theoretical and experimental examination
\cite{Stockmann,Guhr}. However, there are exceptions to this rule, which consist of the special class
of time-reversal invariant systems with point group irreducible representations that can exhibit the
GUE statistics \cite{Leyvraz,Keating}. The family of systems presently known to show this phenomenon is
formed by billiards having threefold symmetry, which have been experimentally implemented
\cite{Dembowski,Schafer} in classical microwave cavities.

%-----------------------
%DESCRIBING THE SYSTEM
%-----------------------

\begin{figure}[h] \includegraphics[width=5.8cm]{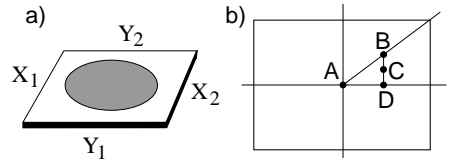}
\caption{
a) Schematic electro-mechanical nanostruture: thin circular quantum
dot on the surface of a suspended dielectric plate.
b) Four different positions for the center of the quantum dot on
the plate surface.}
\label{model}
\end{figure}

In this letter we predict that certain electro-mechanical nanostructures can also exhibit the GOE as
well as the GUE spectral statistics both under TRI conditions. We consider a thin cylindrical quantum
dot (billiard) suspended on a rectangular dielectric nanostructure (phonon cavity), as depicted in Fig.
1a. As the electron-phonon interaction is introduced between the electronic states, restricted to the
circular quantum dot, and the phonon modes of the rectangular cavity an interesting interplay of the
system's two relevant spatial symmetries takes place. In this respect, although having different
dynamics, the present system resembles the much studied Sinai billiard (see, e.g., \cite{Stockmann}),
since in the later case chaos arises from the mismatch between the same two symmetries. It will be
shown that for sufficiently strong electron-phonon coupling, the energy level distribution of this
nanostructure exhibits the GOE  spectral statistics if the center of the quantum dot lies in one of the
symmetry axes of the phonon cavity, whereas the GUE statics occurs if center of the dot is located
outside the symmetry axes (refer to Fig. 1b). We also investigate the influence of material and
geometrical parameters on the unfolding of chaos, indicating the conditions for its experimental
observation.

%---------------
% CALCULATIONS
%---------------

The boundaries of the suspended dielectric plate can be either clamped (C)
or free (F), corresponding to the Dirichlet or Neumann conditions,
respectively.
To obtain the vibrational modes of the plate we use the Classical Plate
Theory (CPT) approximation \cite{Graff}, which is well suited for quasi
two-dimensional mechanical systems. The vibrations of the cavity are thus
described by a vector field whose cartesian components are
\begin{eqnarray}
U(x,y,z,t) = - z \frac{\partial W}{\partial x} \ \ , \ \
V(x,y,z,t) = - z \frac{\partial W}{\partial y}\ , & & \nonumber \\
W(x,y,t) \label{W} = \sum_{m,n} A_{mn} X_m(x)Y_n(y)e^{-i\omega t}.
& &
\label{w(x,y,t)}
\end{eqnarray}
In Eq. (\ref{w(x,y,t)}), $W(x,y,t)$ is written in terms of the one-dimensional transverse modes $X_m(x)
= \sin[k_mx] + \sinh[k_mx] + \zeta_m \left\{\cos[k_mx] + \cosh[k_mx]\right\}$, and likewise for $Y_n$.
The modes $X_m$ and $Y_n$  are the solutions of the Bernoulli-Euler equation \cite{Graff,Leissa} under
the appropriate boundary conditions. The kinetic and strain energies for the cavity are
\begin{eqnarray}
{\cal{K}}(x,y) &=& \rho_{_{2D}} \ \omega^2 W^2(x,y)/2 \ , \\
\label{T_xy}
{\cal{V}}(x,y) &=&
\frac{D}{2} \left\{ \left[\frac{\partial^2 W}{\partial x^2} +
\frac{\partial^2W} {\partial y^2} \right] \right. \\
&-& \left. 2(1-\nu) \left[ \frac{\partial^2 W}{\partial x^2}
\frac{\partial^2 W}{\partial y^2} -
\left(\frac{\partial^2 W}{\partial x \partial y}\right)^2 \right]
\right\}\ , \nonumber \label{V_xy}
\end{eqnarray}
where $\rho_{_{2D}}$, $\nu$ and  $D$ represent the two-dimensional density, the Poisson constant and
the rigidity constant of the material, respectively. The Rayleigh-Ritz method is then used to obtain
the coefficients $A_{ij}$ of Eq. (\ref{w(x,y,t)}), by applying the condition
$\partial\mathcal{U}/\partial A_{ij} = 0$ to the energy functional ${\cal{U}} = \int dx dy
[{\cal{K}}(x,y) - {\cal{V}}(x,y)]$ \cite{Leissa}. Having obtained the vibrational eigenfrequencies
($\omega_{\alpha}$) and the eigenmodes ($A^{\alpha}_{mn}$) of the dielectric cavity, an arbitrary
deflection field is written as ($Q_{\alpha}(t) = Q_{\alpha}\exp[-i\omega t]$)
\begin{equation}
{\bf u} = \sum_{\alpha} \left[Q_{\alpha}(t) +
Q^*_{\alpha}(t)\right] \left[ U_{\alpha}({\bf r}) \, \hat{\imath} +
V_{\alpha}({\bf r}) \, \hat{\jmath} + W_{\alpha}({\bf r}) \, \hat{k}
\right].
\label{modo_geral}
\end{equation}

To provide the same level of description to the elastic and electronic
degrees of freedom of the electro-mechanical nanostructure, we quantize
the deflection field of Eq. (\ref{modo_geral}).
So, we define the  operator $a^{\dagger}$ ($a$) that creates
(annihilates) a mechanical mode as
\begin{equation}
a^{\dagger}_{\alpha}(t) = \sqrt{\frac{V\rho\
\omega_{\alpha}}{2\hbar}}\hat{Q}^{\dagger}_{\alpha}(t) -
i \sqrt{\frac{1}{2\hbar V\rho\ \omega_{\alpha}}}
\hat{P}^{\dagger}_{\alpha}(t).
\end{equation}
Thus the quantum vibrational field that interacts with the electrons in
the circular quantum dot is given by
\begin{equation}
\hat{{\bf u}} =
\sum_{\alpha} \frac{[a_{\alpha}(t) +
a^\dag_{\alpha}(t)]} {\sqrt{2\ V\rho\ \omega_{\alpha}/\hbar}}
\left[ U_{\alpha}({\bf r})\, \hat{\imath} + V_{\alpha}({\bf r})\,
\hat{\jmath} + W_{\alpha}({\bf r})\, \hat{k} \right].
\label{modo_quantico}
\end{equation}

The electrons, in the free electron gas approximation, occupy the
states ($\kappa \equiv (\pm l,\nu)$; $l = 0, 1, 2, \ldots$)
\begin{equation}
\phi_{\kappa}({\bf r}) = \frac{{\mbox J}_l
\left(\alpha_{l\nu}\frac{\rho}{R}\right) \exp[\pm i l \theta]}
{\sqrt{\pi}R|J_{l+1}(\alpha_{l\nu})|}
\sqrt{\frac{2}{d}}\sin{\left(\frac{\pi z}{d}\right)}.
\label{pureelectron}
\end{equation}
Here, $\alpha_{l \nu}$ is the $\nu$-th root of the Bessel function of
order $l$, $R$ is the radius and $d$ the width of the quantum dot.

At low temperatures only the long wavelength acoustic modes are excited
and the phonon cavity can be treated as a continuum elastic medium.
Then, the relevant electron-phonon interactions are due to the deformation
(DP) and piezoelectric (PZ) potentials. The DP coupling operator is
written as $C_{DP}\hat{\Delta}({\bf r})$, where $C_{DP}$ is the
deformation potential constant and $\hat{\Delta}({\bf r})= \nabla
\hat{{\bf u}}({\bf r})$ is the relative volume variation.
The DP electron-phonon Hamiltonian is, therefore,
\begin{eqnarray}
\hat{H}_{DP} &=& C_{DP} \sqrt{\frac{\hbar}{2 V \rho}}
\sum_{\kappa',\alpha,\kappa} \frac{V_{\kappa' \alpha
\kappa}^{DP}}{\sqrt{\omega_{\alpha}}}\ b^\dag_{\kappa'}
\left[a_{\alpha}^\dag + a_{\alpha}\right] b_\kappa\ ,
\label{DP} \\
V_{\kappa' \alpha \kappa}^{DP} &=& \int_{\mathcal{D}}
\phi^*_{\kappa'} \nabla \left(U_{\alpha}, V_{\alpha},
W_{\alpha}\right) \phi_\kappa \ d{\bf r} \nonumber \\
&=& - \sum_{mn} A^{\alpha}_{mn} \int_{\mathcal{D}} z \, \phi^*_{\kappa'}
\left[X''_{m}Y_n + X_m Y''_n \right] \phi_\kappa \ d{\bf r}.
\nonumber
\label{F}
\end{eqnarray}
In Eq. (\ref{DP}),  $b^\dag_\kappa$ ($b_\kappa$) is the electronic
creation (annihilation) operator and the integration is performed on the
domain $\mathcal{D}$, defined by the boundary of the quantum dot billiard.

For a piezoelectric (PZ) nanostructure of cubic crystal symmetry, the electric field produced by the
cavity vibrations is ${\bf E} = \left[ \left(-2 \varrho_{14}/\epsilon\right) \varepsilon_{xy}\right]
\hat{k}$ \cite{Auld}, where $\varrho_{14}$ and $\varepsilon_{xy}$ are elements of the piezoelectric and
strain tensors, respectively, and $\epsilon$ is the dielectric constant. In the CPT approximation
$\varepsilon_{xz}=\varepsilon_{yz}=0$. Thus, the piezoelectric potential is ($\Lambda = d
\left(2z-d\right)/2$)
\begin{eqnarray}
2\frac{\varrho_{14}}{\epsilon}\Lambda(z) \sum_{\alpha} \sqrt{\frac{\hbar}{2V\rho\omega_{\alpha}}} \,
[a_{\alpha} + a^\dag_{\alpha}] \frac{\partial^2 W_{\alpha}}{\partial x \partial y}.
\end{eqnarray}
The PZ electron-phonon Hamiltonian is, therefore,
\begin{eqnarray}
\hat{H}_{PZ} &=& 2 e \frac{\varrho_{14}}{\epsilon}
\sqrt{\frac{\hbar}{2 V \rho}} \sum_{\kappa',\alpha,\kappa}
\frac{V_{\kappa' \alpha \kappa}^{PZ}}{\sqrt{\omega_{\alpha}}} \
b^\dag_{\kappa'} \left[a_{\alpha}^\dag + a_{\alpha}\right] b_\kappa\ ,
\nonumber \\
V_{\kappa' \alpha \kappa}^{PZ} &=&  \sum_{mn} A^{\alpha}_{mn} \int_{\mathcal{D}} \Lambda(z) \
\phi^*_{\kappa'} X'_m Y'_n \phi_\kappa \ d{\bf r}. \label{G}
\end{eqnarray}

The electro-mechanic eigenstates are thus obtained from the diagonalization of Hamiltonian $\hat{H} =
\hat{H}_0 + \hat{H}_{DP} + \hat{H}_{PZ}$ written in the basis $\Big\{|\phi_{\kappa} \rangle
\prod_{\alpha}^N \frac{(a_{\alpha}^\dag)^{n_{\alpha}}}{\sqrt{n_{\alpha}!}} \ |0\rangle\Big\}$ of
eigenstates of $\hat{H}_0 = \sum_\kappa E_\kappa b^\dag_\kappa b_\kappa + \sum_\alpha^N
(\hat{n}_{\alpha} + \frac{1}{2})\hbar\omega_{\alpha}$. $N$ is the number of phonon modes included in
each basis state. It can be verified that $\hat{H}$ is time reversal invariant.

%--------- %RESULTS %---------

We investigate the dynamical behavior of the electro-mechanical
nanostructure by performing a statistical analysis of its spectrum, in
terms of: (i) the eigenenergies nearest neighbor spacing distribution
$P(s)$, measured in units of mean spacing energy; and (ii) the average
spectral rigidity $\overline{\Delta}_3(l)$, which provides information
about the correlation between energy levels within a normalized energy
interval of length $l$ (for technical details see, e.g., Refs.
\cite{Stockmann,Guhr}). For all the analysis we use the first 2500
states, out of 15000 basis states, which suffice to produce very good
spectral statistics.

From exhaustive calculations we found that for sufficiently strong
electron-phonon coupling the observed chaotic behavior of the
nanostructure proved to be quite robust with respect to variations of
geometrical parameters, boundary conditions and basis size.
Thus, for the sake of clearness, throughout this work we consider the
representative case of a quantum dot of  radius $R=450\ nm$ and thickness
$d = \delta/5$ on the surface of a square phonon cavity of sides $L = 1\
\mu m$ and width $\delta = 40\ nm$.
We also assume the clamped (C) boundary conditions for the four edges of the
cavity.
It has been verified that the results are similar for the different boundary
conditions, however the clamped only
($X_1 X_2 Y_1 Y_2 = $\{CCCC\}) plate demonstrates more clearly
the roles of the deformation potential and piezoelectric interactions.
Nonetheless, the variation of some material parameters affects the chaotic
behavior, for instance, the chaos is stronger for softer phonon cavities and
for larger electronic effective masses.
In the calculations we use the material parameters corresponding to an AlAs
dielectric phonon cavity and an Al$_{0.5}$Ga$_{0.5}$As quantum dot.
This choice takes advantage of the very small lattice parameter mismatch
of the interface as well as the large effective mass of the $X$ valley of
AlGaAs \cite{Adachi}.
In addition, to set the necessary strength of the electron-phonon coupling
we multiply the corresponding interaction potentials by an arbitrary factor
$\beta$.
Finally, we notice that the observed chaotic behavior does not result from
the coupling a regular system (the quantum dot) with an already chaotic
system (the phonon cavity). Indeed, by analyzing the spectral statistics of
the vibrating plate alone we find a rather regular behavior.

\begin{figure}[h] \includegraphics[width=7.8cm]{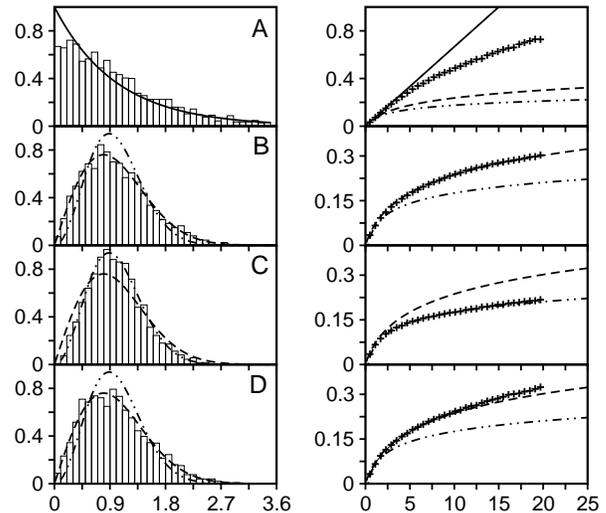}
\caption{
The level statistics for the nanostructure with the DP interaction only
and the quantum dot centered at positions A, B, C and D, Fig. 1b.
The left (right) panel represents by histograms ($+$ symbols) the
numerically calculated $P(s)$ ($\overline{\Delta}_3(l)$) distribution.
The curves indicate the expected behavior for regular (solid), chaotic
GOE-type (dashed), and chaotic GUE-type (dot-dashed) systems.
Here $\beta=10$.}
\label{statistics}
\end{figure}

The interplay between the cylindrical and rectangular symmetries, through
the electron-phonon coupling, destroys the all the geometrical invariances,
except for the reflection symmetries if the center of the quantum dot is
located on a symmetry axis of the plate.
This scenario corresponds to the points A, B and D of Fig. 1b.
On the other hand, for the center of the quantum dot located at C, no
symmetries remain.
A slight displacement of the quantum dot is enough to generate a chaotic
behavior, thus the relative coordinates used in the calculations are:
A = (0,0), B = (0.05,0.05), C = (0.05,0.025), and D = (0.05,0).
Figure \ref{statistics} shows, for the DP interaction only and $\beta=10$,
$P(s)$ and $\overline{\Delta}_3(l)$ for these four cases.
In the case A, the spectral statistics indicates a somewhat regular
dynamics \cite{FFFC}, but in B the occurrence of quantum chaos is clear
and the level distributions are well described by the predictions of GOE
random matrices.
The same occurring for the equivalent case D.
The more interesting result, however, is obtained for C, for which the
statistics belongs to the GUE class, although the system is time-reversal
invariant.

This effect was first predicted by Leyvraz et al. \cite{Leyvraz} and observed experimentally
\cite{Dembowski,Schafer} in microwave billiards with only the threefold symmetry. In such case there
are two classes of eigenstates: real singlets and complex conjugate doublets, which present the GOE and
GUE statistics, respectively. In our case the electron states, Eq. (\ref{pureelectron}), naturally
provide the necessary complex representation through the angular momentum quantum number $l$. Note that
$l=0$ and $l=\pm 1,\pm 2, \ldots$ play, respectively, the role of singlet and doublet states. The
coupling between electron and phonon systems is responsible for generating chaos and, according to
$\hat{H}_{DP}$ and $\hat{H}_{PZ}$, the electron interacts with phonon modes $\alpha$, each of definite
parity. If the dot is centered at a symmetrical locus, e.g., B or D, the parity makes the Hamiltonian
matrix real (up to a global phase) and the GOE statistics is obtained. When the reflection symmetry is
broken, as in C, the electronic angular momenta are coupled and the Hamiltonian is complex, exhibiting
the GUE statistics (a detailed analysis will appear elsewhere \cite{Gusso}). The same effect is
obtained when the boundary conditions of the cavity are changed, which can generate, for instance, the
GUE statistics in locus D for the \{FFFC\} nanostructure.

\begin{figure}[h] \includegraphics[width=5.4cm]{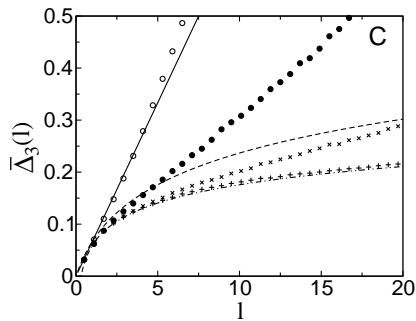}
\caption{
The calculated spectral rigidity for case C and $\beta$ equal to:
1 (open circle), 3 (filled circle), 5 ($\times$) and 10 ($+$).
Once more the curves represent regular (solid), GOE (dashed) and GUE
(dot-dashed) type of systems.}
\label{beta}
\end{figure}

The dependence of the spectral rigidity $\overline{\Delta}_3(l)$ on the
electron-phonon coupling strength is illustrated in Fig. \ref{beta}.
For the situation C and again considering only the DP interaction, we
take $\beta$ = 1, 3, 5, and 10.
As $\beta$ increases, the calculated statistics gradually converge to the
GUE prediction.
Notice that the numerical results are never well fitted by the GOE case.
The inclusion of more basis states does not change the observed results
significantly, however, the effect is favored by changing some material
parameters, for instance, by using softer phonon cavities, thiner plates,
and including the PZ interaction.

\begin{figure} \includegraphics[width=5.4cm]{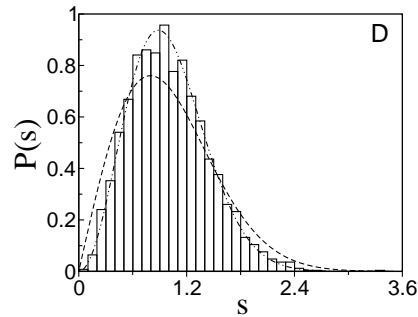}
\caption{The nearest neighbor spacing distribution for case D, where the
DP and PZ interactions are included and $\beta=10$ for both.
The curves correspond to GOE (dashed) and  GUE (dot-dashed) statistics.}
\label{PZ}
\end{figure}

Finally, examination of equations (\ref{F}) and (\ref{G}) shows that,
individually, both the DP and PZ interactions preserve the reflection
symmetry of the matrix elements with respect to the \{x,y\} axes and the
diagonal axes.
However, when acting together the reflection invariance is broken and the
spectrum statistics of loci B and D must change from GOE to GUE.
Fig. \ref{PZ} confirms this effect by showing the $P(s)$ distribution for
the quantum dot at D, with the DP as well as PZ interactions included with
$\beta=$ 10. The agreement with the GUE statistics is excellent, in
contrast to case D of Fig. \ref{statistics}.
Because the AlGaAs alloy is a weak piezoelectric material, the DP coupling
shows a stronger effect in promoting the chaos, whereas the main action of
the PZ interaction (in the presence of DP) is to break the system's
spacial symmetry.

%---------------- % FINAL REMARKS %----------------

To conclude we  briefly describe results obtained for other types of suspended structures, which will
be included in a forthcoming publication \cite{Gusso}. Regarding the phonon cavities, we investigated
different boundary conditions, in particular \{FFFF\} and \{CFCF\}. The conclusions are essentially the
same and the ideas presented here can be extrapolated to these distinct cases. We also considered a
quantum dot of rectangular symmetry. In such case chaotic behavior was observed only for very
particular asymmetric configurations, but never resulting in GUE statistics. This aspect shows the
importance of the interplay between the circular symmetry of the electronic states with the rectangular
symmetry of the cavity phonon modes and, therefore, the fundamental relevance of the architecture of
the nanostructures.

CNPq/CT-Energ and Finepe/CT-Infra1 (M.L.), and CNPq (A.G., M.L.) are acknowledge for research grants.

\end{document}